\documentclass[english]{article}
\usepackage{mathpazo}
\usepackage[latin9]{inputenc}
\usepackage[letterpaper]{geometry}
\usepackage{babel}
\usepackage{float}
\usepackage{textcomp}
\usepackage{amsmath}
\usepackage{amssymb}
\usepackage{graphicx}
\usepackage{setspace}
\PassOptionsToPackage{normalem}{ulem}
\usepackage{ulem}
\usepackage{microtype}
\usepackage[unicode=true,
 bookmarks=true,bookmarksnumbered=false,bookmarksopen=false,
 breaklinks=false,pdfborder={0 0 1},backref=false,colorlinks=false]
 {hyperref}

\makeatletter
\newcommand{\lyxaddress}[1]{
	\par {\raggedright #1
	\vspace{1.4em}
	\noindent\par}
}

\@ifundefined{date}{}{\date{}}

\usepackage[margin=1cm]{caption}

\renewcommand\[{\begin{equation}}
\renewcommand\]{\end{equation}}

\usepackage{fontawesome5}

\usepackage{tocloft}
\makeatother

\begin{document}
\global\long\def\d{\mathrm{d}}%
\global\long\def\sp{,\qquad}%

\title{Wormhole Time Machines and Multiple Histories}
\author{\textbf{Barak Shoshany}{\small{}}\\
{\small{}\faIcon{envelope} \href{mailto:bshoshany@brocku.ca}{bshoshany@brocku.ca}$\qquad$\faIcon{globe}
\href{https://baraksh.com/}{https://baraksh.com/}}\\
{\small{}\faIcon{orcid} \href{https://orcid.org/0000-0003-2222-127X}{0000-0003-2222-127X}$\qquad$\faIcon{github}
\href{https://github.com/bshoshany}{@bshoshany}}\\
\\
\textbf{Jared Wogan}\\
{\small{}\faIcon{envelope} \href{mailto:jared.wogan@gmail.com}{jared.wogan@gmail.com}$\qquad$\faIcon{globe}
\href{https://jaredwogan.ca/}{https://jaredwogan.ca/}}\\
{\small{}\faIcon{orcid} \href{https://orcid.org/0000-0002-0737-5492}{0000-0002-0737-5492}$\qquad$\faIcon{github}
\href{https://github.com/JaredWogan}{@JaredWogan}}}
\maketitle

\lyxaddress{\begin{center}
\faIcon{university} \href{https://brocku.ca/}{Department of Physics, Brock University}\\
\faIcon{map-marker-alt} \href{https://goo.gl/maps/qscBMigohESxxczM7}{1812 Sir Isaac Brock Way, St. Catharines, Ontario, L2S 3A1, Canada}
\par\end{center}}
\begin{abstract}
In a previous paper, we showed that a class of time travel paradoxes
which cannot be resolved using Novikov's self-consistency conjecture
can be resolved by assuming the existence of multiple histories or
parallel timelines. However, our proof was obtained using a simplistic
toy model, which was formulated using contrived laws of physics. In
the present paper we define and analyze a new model of time travel
paradoxes, which is more compatible with known physics. This model
consists of a traversable Morris-Thorne wormhole time machine in 3+1
spacetime dimensions. We define the spacetime topology and geometry
of the model, calculate the geodesics of objects passing through the
time machine, and prove that this model inevitably leads to paradoxes
which cannot be resolved using Novikov's conjecture, but can be resolved
using multiple histories. An open-source simulation of our new model
using Mathematica is available for download on GitHub. We also provide
additional arguments against the Novikov self-consistency conjecture
by considering two new paradoxes, the switch paradox and the password
paradox, for which assuming self-consistency inevitably leads to counter-intuitive
consequences. Our new results provide more substantial support to
our claim that if time travel is possible, then multiple histories
or parallel timelines must also be possible.
\end{abstract}
\newpage{}

\tableofcontents{}

\section{Introduction}

\subsection{\label{subsec:Time-travel}Time travel and its paradoxes}

Einstein's theory of general relativity has been around for more than
100 years, and has been verified experimentally to very high accuracy.
However, much about this theory is still not well-understood. One
particularly interesting question that has so far remained unanswered
is whether it is possible, within this theory, to violate causality
by traveling back in time.

Indeed, there are many spacetime geometries within general relativity
\cite{Shoshany_FTL_TT,Krasnikov,Lobo,EverettRoman} which seem to
allow time travel, defined more precisely by the existence of \textbf{closed
causal (timelike or null) curves}. In this paper we will focus on
one such example: a \textbf{wormhole} \cite{MorrisThorne88,visser}.

A wormhole can be roughly defined as a ``shortcut'' from one point
in spacetime to another, allowing a traveler to go from point A directly
to point B without traversing the distance between them. If a wormhole
can connect two points in space, then it stands to reason that it
could perhaps also connect two different points in time, thus forming
a time machine.

In fact, even if the wormhole originally only connected two points
in space which correspond to the same moment in time, one might be
able to use relativistic time dilation to make time at one mouth pass
slower than time in the other mouth, thus converting it into a time
machine \cite{PhysRevLett.61.1446}.

But if time travel is possible, it could lead to paradoxes \cite{Krasnikov02,Krasnikov97,wasserman2018paradoxes}.
This includes two main types of paradoxes:
\begin{itemize}
\item \textbf{Consistency paradoxes}, where the chain of events is inconsistent.
For example, imagine that Alice goes into her time machine, travels
back 5 minutes to the past, and then destroys the time machine. Since
the machine is now destroyed, Alice won't be able to travel back and
destroy the machine. But this means the machine will be in working
order, so Alice \textbf{will }be able to travel back in time and destroy
it. In other words, the machine exists if and only if it is destroyed,
leading to an inconsistency.
\item \textbf{Bootstrap paradoxes}, where an event causes itself, or something
is created out of nothing. For example, imagine that Bob receives
plans for a time machine from his future self, who arrives in a similar
time machine. Bob then builds the machine, goes back in time, and
gives himself the plans. Everything is perfectly consistent, so this
is not a consistency paradox. However, Bob's construction of the time
machine causes itself to happen, and the plans for the time machine
were created from nothing\footnote{We will discuss in more detail why this is considered a paradox in
section \ref{subsec:Bootstrap-paradoxes}.}.
\end{itemize}
How are these paradoxes resolved? One option is to assume that time
travel is simply impossible in the first place, as \textbf{Hawking's
chronology protection conjecture} claims \cite{Hawking92,visser1993wormhole,visser2003quantum}.
However, at the moment, this remains merely an unproven conjecture
\cite{earman2009laws}. Indeed, a conclusive proof of this conjecture
would require a consistent and experimentally-proven theory of quantum
gravity, which does not yet exist.

Still, there have been some attempts to prove this conjecture using
quantum field theory on curved spacetime and semi-classical gravity.
For example, in \cite{KimThorne} it was shown that the renormalized
energy-momentum tensor of a certain quantum field diverges in the
presence of a time machine. However, the authors commented that such
divergences should get cut off by quantum gravity effects, and \cite{Krasnikov1996}
later presented spacetimes containing time machines where the energy-momentum
tensor in fact does not diverge.

Similarly, \cite{kay1997quantum} introduced theorems which show that
the renormalized expectation value of a quantum scalar field and its
energy-momentum tensor are ill-defined or diverge in the presence
of time machines, but \cite{krasnikov1998quantum} later showed that
this can be avoided by replacing some assumptions. We thus see that
Hawking's conjecture is, at present, still far from being proven,
and therefore time travel remains a possibility.

If time travel is possible, then it must be possible without paradoxes.
One certainly cannot imagine a universe where the time machine is
simultaneously both destroyed and not destroyed\footnote{One may wonder if perhaps this can be achieved by allowing the time
machine to be in a \textbf{superposition }of destroyed and not destroyed.
Indeed, we will discuss how quantum mechanics, in the context of the
Everett (``many-worlds'') interpretation, can be used to resolve
time travel paradoxes in section \ref{subsec:The-nature-of}.}. The simplest way to avoid paradoxes, while still allowing for time
travel to occur, is via the \textbf{Novikov}\footnote{The Novikov self-consistency conjecture (sometimes also called the
Novikov self-consistency principle) is named after physicist Igor
Novikov, and should not be confused with another ``Novikov conjecture'',
named for mathematician Sergei Novikov, which is related to topology.}\textbf{ self-consistency conjecture} \cite{PhysRevD.42.1915}, which
suggests that one can simply never make any changes to the past.

According to this conjecture, any attempts to change the past will
necessarily either fail, or bring about the very future they tried
to prevent. If the past cannot be changed, then there is also no possibility
of paradoxes, no matter how many times you travel to the past.

Bootstrap paradoxes, however, still remain even if Novikov's conjecture
is correct, as we will discuss in section \ref{subsec:Bootstrap-paradoxes}.
For example, the bootstrap paradox scenario we described above is
100\% compatible with Novikov's conjecture, as is has no inconsistencies.

\subsection{Resolving paradoxes with multiple histories}

It has been shown \cite{Consortium91} that Novikov's conjecture can
be applied to certain simple consistency paradoxes. However, in a
previous paper \cite{ParadoxModel}, we\footnote{More precisely, one of us, Barak Shoshany, along with his student
Jacob Hauser.} analyzed a physical system for which it was previously proven \cite{Krasnikov02}
that the Novikov conjecture cannot apply under any circumstances.

In the same paper, we also showed that the same paradox that cannot
be resolved using Novikov can, in fact, be completely resolved by
assuming the existence of \textbf{multiple histories}, or equivalently,
\textbf{parallel timelines}. Since the histories are independent,
the paradoxical chain of events breaks:
\begin{itemize}
\item Alice steps into the time machine in history 1, but emerges from it
in history 2.
\item When Alice destroys the time machine in history 2, there is no inconsistency,
as this is not the same machine she used to travel back in time, which
is still intact -- in history 1.
\end{itemize}
Furthermore, unlike Novikov's conjecture, multiple histories can also
resolve bootstrap paradoxes. For example, the bootstrap paradox described
above is paradoxical because Bob gave himself the plans for the time
machine, but never created the plans in the first place. This paradox
may easily be resolved by considering that in this scenario, there
are in fact \textbf{three }Bobs in three separate histories:
\begin{itemize}
\item Bob 1 creates the plans for the time machine in history 1, then travels
back in time, emerges in history 2, and gives Bob 2 the plans.
\item Bob 2 then builds the machine, travels back in time, emerges in history
3, and gives Bob 3 the plans.
\end{itemize}
While neither Bob 2 nor Bob 3 created the plans, we now see that the
plans were, in fact, created by Bob 1, so there is no paradox here.

\subsection{\label{subsec:Our-old-and}Our old and new paradox models}

The idea of resolving time travel paradoxes using multiple histories
is, of course, not new; see for example the discussion in \cite{visser}
(chapter 19), \cite{Everett:2004ka}, \cite{Deutsch91}, and \cite{deutsch1994quantum}.
However, to our knowledge, this idea remained strictly abstract for
a long time, with no concrete examples of explicit paradoxes resolved
by multiple histories presented in the literature -- until the publication
of our previous paper, \cite{ParadoxModel}, where we introduced such
a concrete example for the first time. However, our previous model
was not without issues.

The main problem (and source of criticism \cite{WHYTE20196}) of our
previous paper is that the physical model we used to analyze the time
travel paradoxes was a very simplistic toy model. Therefore, one might
argue that perhaps this model cannot be used to prove results about
the real world, so maybe our conclusions were invalid, and Novikov's
conjecture is safe after all. In the present paper, we remedy this
problem by performing a similar analysis, and reproducing the same
proof, with a more realistic model, as detailed in section \ref{subsec:The-new-and}.

Of course, we do not yet know if wormholes can exist in reality --
and even if they do, we do not yet know if they can be used for time
travel. However, the new model is completely ``realistic'' in the
sense that if (traversable) wormholes \textbf{did} exist, and \textbf{could}
be used for time travel, then the theory of general relativity, as
we understand it today, appears to predict that one could in principle,
using sufficiently advanced technology, construct the physical system
described by our model and perform real experiments with it, and its
behavior will be governed by familiar and well-tested laws of physics.

In this paper, we prove that this new and more realistic model inevitably
leads to paradoxes which cannot be resolved using Novikov's conjecture,
but can be resolved using multiple histories. We thus further strengthen
our claim that time travel necessarily implies multiple histories.
If multiple histories do \textbf{not }exist, then time travel would
inevitably lead to paradoxes, and thus to inconsistent physics, which
would make it impossible.

\section{In-depth comparison of the old and new models}

\subsection{\label{subsec:The-original-toy}The original toy model}

Let us recall the toy model used in the previous paper \cite{ParadoxModel},
which was based on a previous model by Krasnikov \cite{Krasnikov02}.
The model is formulated in the \textbf{twisted Deutsch-Politzer (TDP)}
space in 1+1 spacetime dimensions, with coordinates $(t,x)$. This
space is constructed by associating the interval $(1,x)$ with $(-1,-x)$
for $-1<x<1$. TDP space is meant to be a toy model of a wormhole
time machine, but as we will see, the two are not quite analogous.

A particle entering the interval at $t=1$ will travel back in time
to $t=-1$. Furthermore, it will be ``twisted'', with its spatial
orientation inverted. The reason for the twisting is to ensure that
the particle emerging at $t=-1$ will collide with its past self and
potentially create a paradox; if both particles moved in the same
direction, they would never collide.

To obtain time travel paradoxes in this model, we impose a few simple
physical laws:
\begin{enumerate}
\item The particles are all \textbf{massless}, and move along lightlike
(or null) paths.
\item Each particle can have one of two \textbf{colors}, for example blue
and green\footnote{In \cite{ParadoxModel} we also analyzed the case where additional
colors and/or particles are allowed, but this will not be relevant
to the current discussion.}.
\item Whenever two particle worldlines intersect, the particles interact.
Each particle flips both its \textbf{direction of motion} and its
\textbf{color}, independently of the color of the other particle.
A blue particle turns green and a green particle turns blue.
\end{enumerate}
The last law is crucial, as without it, time travel paradoxes could
be easily avoided using Novikov's conjecture \cite{Consortium91,Krasnikov02}.
By allowing the particles to have colors, which make them \textbf{distinct}
from one another, and by having them interact in this particular way,
we ensure that the there cannot be a consistent evolution -- and
a paradox is always created.

\begin{figure}[H]
\centering{}\includegraphics[width=0.66\textwidth]{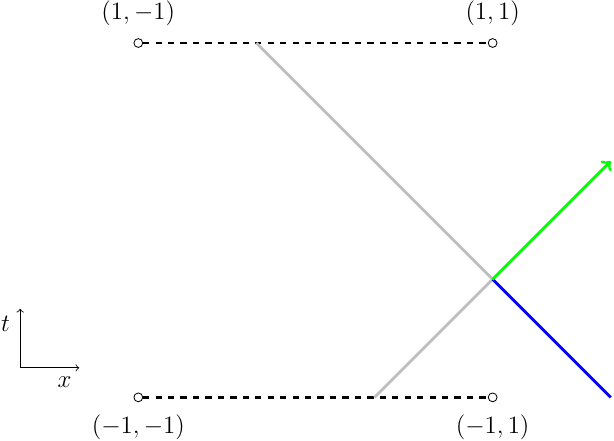}\caption{\label{fig:TDPexample}Example of a paradox in the toy model.}
\end{figure}

For example, in Figure \ref{fig:TDPexample}, a blue particle is approaching
the system from the right, and collides with its future self. This
collision results in the two colliding particles switching colors.
The particle that emerged from the time machine then continues, enters
the time machine at $t=1$, and exits at $t=-1$. If the particles
did not have colors, then this would be perfectly consistent, and
we could have said that Novikov's conjecture applies to this model.

However, since the particles do have colors, and since the particle
that goes into the time machine at $t=1$ is the \textbf{same }particle
that comes out of it at $t=-1$, the color of the two gray lines in
the figure must be the same. The reader should verify that \textbf{there
is no consistent choice of color} for the two gray lines, since our
imposed physical laws force both particles to switch colors in the
collision.

In other words, the two gray lines must have the \textbf{same }color
since they are the same particle, but they must also have \textbf{different}
colors due to the imposed interaction vertex. Therefore, we have a
consistency paradox which cannot be resolved by Novikov's conjecture.

This result is independent of the initial conditions; the particle
can come from any point in space, and can have any initial color.
As long as it eventually goes into the time machine, there is a paradox.
This means that Novikov's conjecture cannot be applied to this model
\textbf{under any circumstances}, and we must resolve the paradox
in another way.

\begin{figure}[H]
\centering{}\includegraphics[width=1\textwidth]{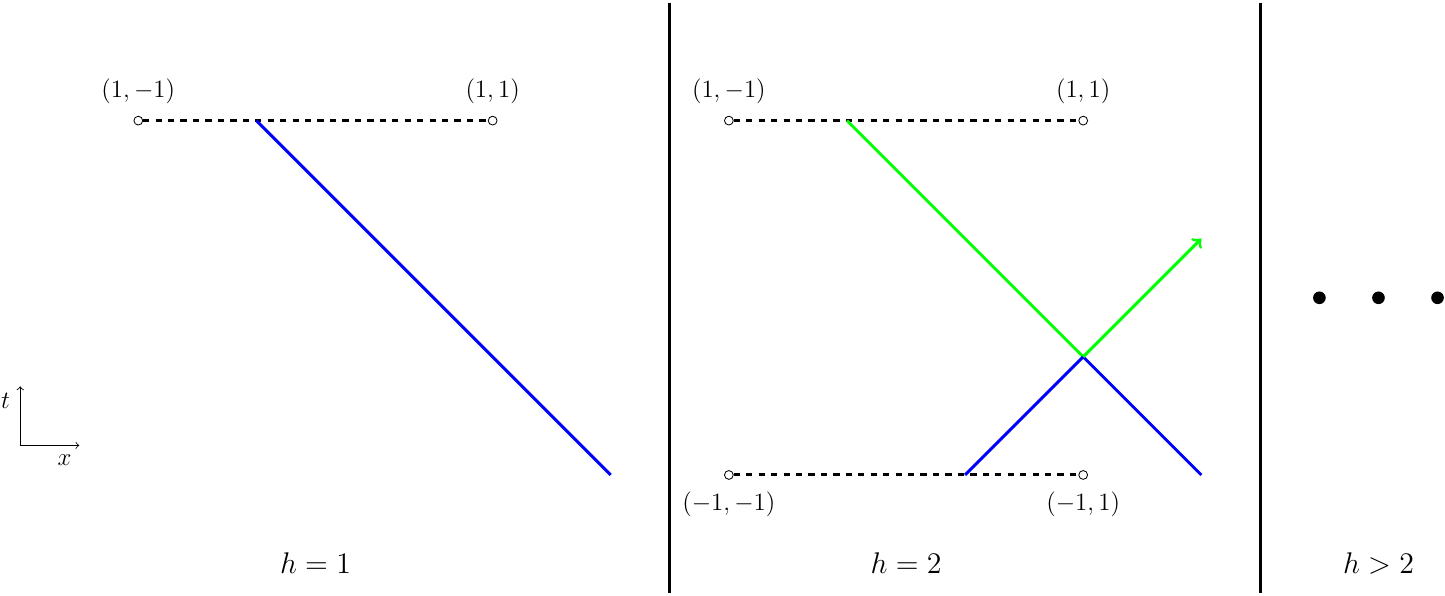}\caption{\label{fig:Resolving-a-paradox}Resolving a paradox in the toy model
using multiple histories.}
\end{figure}

In \cite{ParadoxModel}, we proved that multiple histories provide
a suitable resolution to the paradoxes for all possible initial conditions.
This is illustrated in Figure \ref{fig:Resolving-a-paradox}. The
blue particle starts in history $h=1$ and goes into the time machine
undisturbed; it did not ``yet'' travel through the time machine,
so there is no future particle for it to collide with.

In history 2, the path of the blue particle is the same up to the
point where it collides with its copy from history 1. At that point,
there is an interaction, and the particle is prevented from going
into the time machine. Instead, the particle from history 1, which
has now turned green, goes into the time machine.

The color that goes into the time machine is now \textbf{different
}from the color that came out of it, but that is fine, because the
particle that goes in is \textbf{not }the same particle that came
out. Thus, there is no paradox. Note that this requires an infinite
number of histories, since this process will continue indefinitely.
In \cite{ParadoxModel} we considered some other scenarios which only
require a finite number of histories, but we will not repeat that
discussion here.

Finally, we note that, in the case where there are no colors, Novikov's
conjecture can resolve the consistency paradox, but not the bootstrap
paradox. Since the particles collide elastically, the particle that
comes out of the time machine is the same one that goes into it, so
it has no point of origin -- it only exists without the closed causal
curve, and was essentially created from nothing. However, in the case
of multiple histories, the particle that comes out of the time machine
in history 2 originated from history 1, so there is no bootstrap paradox.

\subsection{\label{subsec:The-new-and}The new and improved model}

In this paper, we will present a more realistic paradox model using
the \textbf{Morris-Thorne traversable wormhole metric} \cite{MorrisThorne88}.
The main differences between the two models were summarized in section
\ref{subsec:Our-old-and}. Let us go over these differences in more
detail now.

\subsubsection*{1+1 vs. 3+1 dimensions}

The toy model was formulated in 1+1 spacetime dimensions. This greatly
simplified the analysis, but the real universe has 2 additional spatial
dimensions that are unaccounted for. Furthermore, in 1+1 dimensions,
general relativity is trivial (as the Einstein tensor is identically
zero), so that the toy spacetime is not a true general-relativistic
spacetime. By increasing the number of spacetime dimensions to 3+1,
we make the model more realistic and permit studying the effects of
gravity.

We will use spherical coordinates, with the objects moving only along
radial geodesics. This means that effectively, the objects are still
moving in a 1+1-dimensional hypersurface -- but that is to be expected,
as there is no reason for them to make turns at any point. While the
model could be further generalized by allowing non-radial geodesics,
that would merely complicate the model without adding any further
insights into time travel paradoxes.

\subsubsection*{Color vs. temperature}

The particle ``colors'' used in the toy model can be thought of
as a discrete property similar to electric charge or QCD color charge.
In this sense, the colors themselves are not unrealistic. However,
the interaction vertex where each particle simply flips its color
in each collision is artificial, and does not have an analogue in
any known laws of physics; in fact, if the colors are charges, then
this vertex clearly violates conservation of charge.

In the new model, we replace color with temperature. In thermodynamics,
two systems in mutual contact will exchange heat until they reach
thermodynamic equilibrium, with heat flowing from the hotter object
to the colder one. Thus, we replace the particles with objects that
have temperature, and assume that the environment is sufficiently
cold that the objects will continuously lose heat over a sufficiently
long amount of time.

By replacing particles and colors with objects and temperatures, we
exchange the contrived physical laws of the toy model with the well-established
laws of thermodynamics. As we will see below, the gradually decreasing
temperature provides essentially the same mechanism for generating
inconsistencies that the colors provided in the toy model.

\subsubsection*{Massless vs. massive}

In the toy model, we only considered massless particles moving along
lightlike (null) trajectories. This made the analysis easier, as the
particles only moved in $\text{45°}$ angles in the spacetime diagrams.
However, it also severely limited the applicability of the model,
as one usually wants to send massive objects through time machines,
not just photons.

In the new model we remedy this by considering massive objects moving
along timelike trajectories, with any speed $v<1$ (in units where
$c\equiv1$). Massless ``objects'' can also be considered, either
by taking the limit as $m\to0$ and $v\to1$, or by taking the ``object''
to be a gas of photons. In this way, we ensure that the new model
is as general as possible and can handle both massive and massless
objects.

\subsubsection*{Flat hole vs. wormhole}

The TDP space used in the toy model is a 1+1-dimensional flat spacetime
with a ``hole''. There is no geometry, only topology, so nothing
interesting is happening in terms of gravity. The entire analysis
is performed using special relativity -- in fact, essentially using
just Newtonian mechanics -- and particles do not experience any gravity.

The TDP time machine itself is an idealized one -- merely a topological
identification of two lines, without any physical structure. The TDP
space also has another problem which, for simplicity, we chose to
ignore in our previous paper: the four points at $\left(t,x\right)=\left(\pm1,\pm1\right)$
are \textbf{singularities}, and this turns out to have some bizarre
consequences \cite[section 3.3]{Krasnikov}.

In the new model, we instead construct a time machine using a physical
wormhole. This means that we are no longer using a purely topological
time machine, but a fully geometric one, where the effects of gravity
can be explored. This allows us to accurately describe how objects
would realistically move in this spacetime if it was possible to construct
it, by considering solutions to the geodesic equations.

The wormhole time machine we will explore exists eternally, and sends
any object that enters it at any time a fixed duration back in time.
This is in contrast to the TDP space time machine, where the wormhole
only comes into existence at two particular points in time, sends
objects from the future point to the past point, and does not exist
at any other times. As we will show, the new model still results in
inevitable paradoxes that can only be removed by assuming multiple
histories.

\section{The wormhole time machine}

\subsection{The metric}

It is now finally time to introduce the mathematical details of the
new model. Let us consider a simplified version of the (static and
spherically symmetric) \textbf{Morris-Thorne traversable wormhole
metric} \cite{MorrisThorne88,visser}:
\begin{equation}
\d s^{2}=-\d t^{2}+\d l^{2}+(b_{0}^{2}+l^{2})(\d\theta^{2}+\sin^{2}\theta\thinspace\d\phi^{2}),\label{eq:metric}
\end{equation}
where $t\in\mathbb{R}$, $l\in\mathbb{R}$, $\theta\in\left[0,\pi\right]$,
and $\phi\in\left[0,2\pi\right)$ are the coordinates, and $b_{0}\in\left[0,\infty\right)$
is a constant. The coordinate $l$ acts as a radial coordinate, except
that it can also be negative. The positive range corresponds to one
region, and the negative range corresponds to another region. Both
regions are asymptotically flat in the limit $|l|\rightarrow\infty$,
and they are connected at the point $l=0$, called the \textbf{throat
}of the wormhole. The metric is symmetric under $l\mapsto-l$.

Objects can travel from one region, through the wormhole's throat,
into the other region. The choice of regions is up to us. They can
be two completely separate universes, or they can be two parts of
the same\textbf{ }universe. In the latter case, the wormhole can be
a shortcut\textbf{ }through space, connecting two very distant locations
and thus potentially allowing faster-than-light travel.

\begin{figure}[h]
\centering{}\includegraphics[width=1\textwidth]{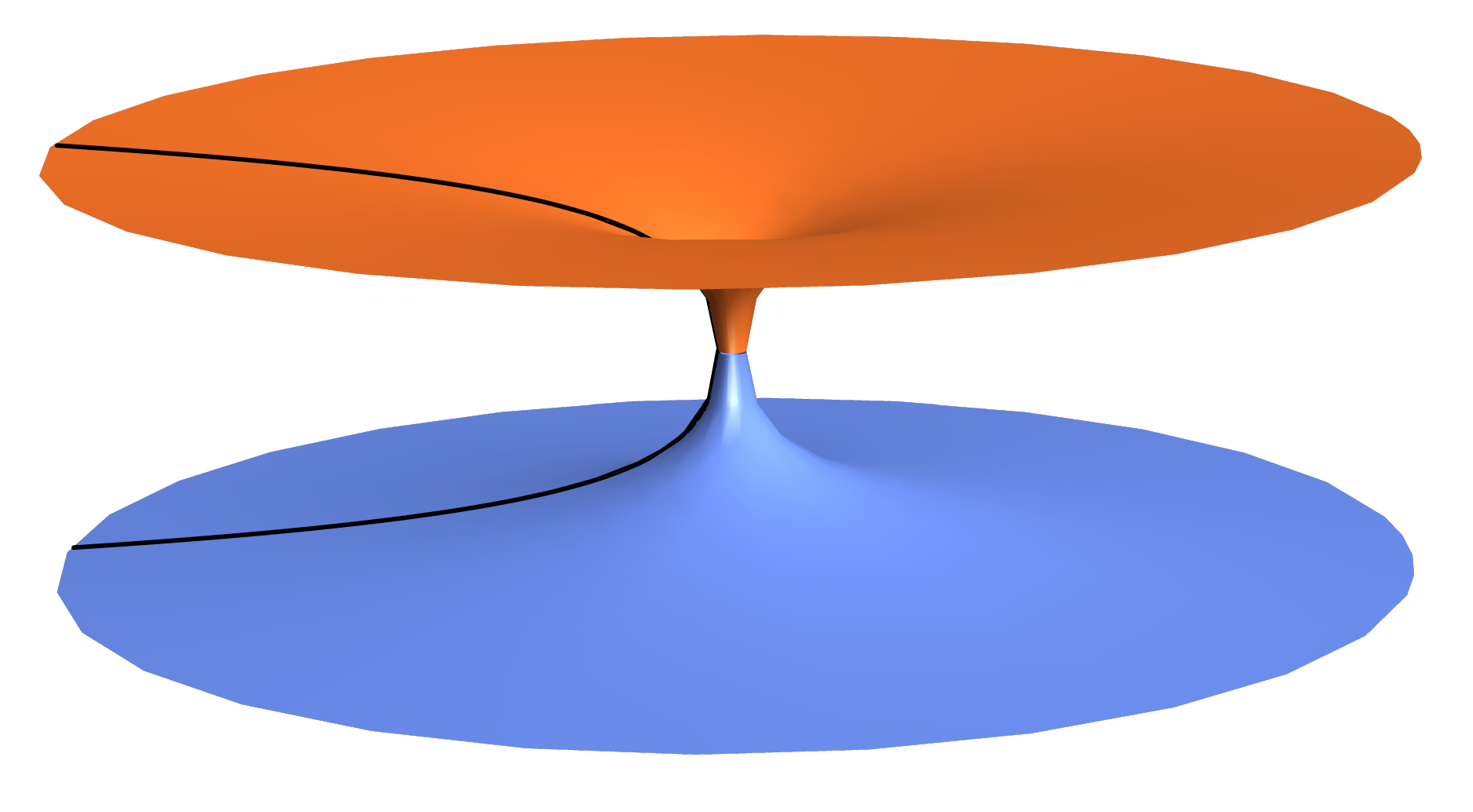}\caption{\label{fig:embeddingWormhole}Embedding diagram of the Morris-Thorne
wormhole metric with $\theta=\pi/2$. Each circle represents a slice
of a sphere with radius $r$. The orange surface represents positive
$l$, while the blue surface represents negative $l$. A possible
geodesic, crossing through the wormhole from one region to the other,
is highlighted in black in the diagram.}
\end{figure}

We can also introduce a different radial coordinate $r$ defined by
$r^{2}\equiv b_{0}^{2}+l^{2}$. With this coordinate, the metric takes
the form
\[
\d s^{2}=-\d t^{2}+\cfrac{\d r^{2}}{1-b_{0}^{2}/r^{2}}+r^{2}\left(\d\theta^{2}+\sin^{2}\theta\thinspace\d\phi^{2}\right).
\]
Note that $r$ is strictly positive, and the smallest value it can
have is $b_{0}$ (corresponding to $l=0$), which means the region
$\left[0,b_{0}\right)$ is inaccessible. $b_{0}$ represents the radius
of the wormhole throat. Furthermore, while $l$ covers both regions,
$r$ only covers one region, so copies of the coordinate system $\left(t,r,\theta,\phi\right)$
are needed to cover the entire wormhole. The embedding diagram for
the wormhole (see \cite{MorrisThorne88} for derivation) can be seen
in Figure \ref{fig:embeddingWormhole}.

Morris-Thorne wormholes have been studied extensively in the literature,
as summarized in the books \cite{visser}, \cite{Lobo}, and \cite{Krasnikov}.
One interesting question that remains unanswered is whether such wormholes
are stable under perturbations, such as those produced due to sending
matter through the wormhole.

The stability of the wormhole largely depends on the properties of
the matter used to construct it. Solving the Einstein equation for
the Morris-Thorne wormhole metric (\ref{eq:metric}) suggests that
it must be supported by ``exotic matter'', which violates some of
the energy conditions \cite{PhysRevLett.61.1446,VisserBarcelo,Curiel:2014zba}.
It is currently unknown whether such matter exists in our universe
in sufficient quantities to build a wormhole.

For example, Armendáriz-Picón \cite{Armendariz-Picon:2002gjc} considered
exotic matter in the form of a minimally-coupled massless scalar field
with a reversed-sign kinetic term, and concluded that wormholes constructed
using it are indeed stable under arbitrary perturbations. However,
it is unclear whether such a scalar field can exist in nature.

More recently, some authors have analyzed the stability of wormholes
using modified theories of gravity. For example, Kuhfittig \cite{Kuhfittig:2020fue}
showed that Morris-Thorne wormholes with zero tidal forces are in
stable equilibrium, as defined by the Tolman-Oppenheimer-Volkov (TOV)
equation. That analysis was performed in $f\left(R\right)$ gravity
with $f\left(R\right)\sim R^{1\pm\epsilon}$, which reproduces Einstein
gravity in the limit $\epsilon\to0$, but it is unclear whether the
results still hold in that limit.

Undoubtedly, much more work is needed in order to determine whether
Morris-Thorne wormholes can be stable in pure Einstein gravity. However,
for the purposes of the time machine model presented in this paper,
we will be working under the assumption that the wormhole is indeed
stable.

\subsection{The geodesics}

In this paper we will only be interested in radial geodesics; allowing
for non-radial geodesics will complicate the calculations while not
adding any new insights into time travel paradoxes, which is our main
goal in the paper. As we are only considering geodesics along $l$,
we have $\d\theta=\d\phi=0$, and the metric simplifies to
\[
\d s^{2}=-\d t^{2}+\d l^{2},
\]
which is a simple 1+1-dimensional flat metric. Thus we immediately
see that the geodesic equation is given by
\[
\ddot{l}\left(t\right)=0.
\]
This equation has the solution
\[
l\left(t\right)=l_{0}+vt,
\]
where $v\in\left[0,1\right]$ is the object's velocity and $l_{0}$
is the initial position at $t=0$. Although the geodesics are just
straight lines in the $l$ coordinate, they become more complicated
when considering the $r$ coordinate, as we will see in the simulations
in section \ref{sec:Simulation-of-the}.

\subsection{Converting the wormhole to a time machine}

Let us now assume that both sides of the wormhole, positive $l$ and
negative $l$, are not only in the same universe, but in fact in the
exact same position, chosen without loss of generality to be the spatial
origin. To convert the wormhole into a time machine, we simply assume
that the origin of one region is shifted by 2 along the $t$ axis
with respect to the origin of the other region. This then means that
any object entering the wormhole in the future region, at time $t$,
will exit it in the past region, at time $t-2$.

There is a slight complication here, as the future and past wormholes
are located on top of each other, so it is unclear if objects entering
the spatial origin are supposed to be sent 2 time units to the past
or to the future. However, we can simply assume that incoming objects
always start at positive $l$, and that negative $l$ is in the past,
so that objects are always sent to the past.

The last property we need to define is the relative spatial orientation
of the two regions, that is, how the direction of the object entering
the wormhole in the future is related to that of the object exiting
the wormhole in the past. Recall from section \ref{subsec:The-original-toy}
that in the TDP space, we had to twist the particle as it went through
the time machine, so that when it comes out it goes back the way it
came from, and is thus guaranteed to collide with itself and create
a paradox.

To reproduce this behavior in the wormhole model, we simply place
the two mouths of the wormhole such that they both have the same orientation.
Both the positive $l$ values and the negative $l$ values are now
mapped to the same spatial coordinate patch (rather than two separate
coordinate patches, as we assumed before), except shifted in time
by 2 units.

Importantly, the $\theta$ and $\phi$ coordinates remain unchanged
after passing through the wormhole. This means that objects coming
in from any radial direction with \textbf{decreasing} $r$ (and $l$
decreasing from $+\infty$ to $0$) will come out with \textbf{increasing}
$r$ (and $l$ decreasing from $0$ to $-\infty$) in the same direction,
and are thus guaranteed to create a paradox by colliding with themselves.

In other words, we are restricting both sides of the wormhole to be
in the same 1+1-dimensional subsurface of the entire 3+1-dimensional
spacetime; this also substantially simplifies the following analysis.

\section{Creating time travel paradoxes\label{sec:Creating-time-travel}}

\subsection{Consistency paradoxes}

The crucial component in creating time travel paradoxes that cannot
be resolved by Novikov's conjecture is introducing a \textbf{distinguishing
property }that all objects going through the time machine have, such
that this property will end up having different values at the entrance
and exit of the time machine, resulting in an inconsistent evolution
and thus an inevitable paradox.

In the toy model, this property was the particle's \textbf{color}
-- an artificial property invented specifically for this model. In
the new wormhole model, we instead use a property that any macroscopic
object already has: \textbf{temperature}. We assume that the objects
are hot and that the environment is cold enough to ensure that the
objects transfer heat to the environment continuously for the entire
duration of the experiment.

More precisely, for a massive object, we assume that it is a black
body satisfying the Stefan-Boltzmann law for radiative cooling:

\begin{equation}
P=\frac{\d E}{\d\tau}=\varepsilon\sigma A\left(T^{4}-T_{\mathrm{ambient}}^{4}\right),
\end{equation}

where $P$ is the total power radiated from the object, $E$ is the
energy of the object, $\tau$ is time, $\varepsilon$ is the emissivity
($\epsilon=1$ for a black body), $\sigma$ is the Stefan-Boltzmann
constant\footnote{Defined since 2019 to have the exact value $\sigma\equiv2\pi^{5}k^{4}/15c^{2}h^{3}$
where $k$ is the Boltzmann constant, $c$ is the speed of light,
and $h$ is the Planck constant.}, $A$ is the surface area of the object, $T$ is the temperature
of the object, and $T_{\mathrm{ambient}}$ is the temperature of the
environment. The specific heat capacity $c$ of the object is defined
by the relation 
\[
c=\frac{1}{m}\frac{\d E}{\d T}\qquad\Rightarrow\qquad mc=\frac{\d E}{\d T},
\]
where $m$ is the object's mass. Multiplying by $\d T/\d\tau$ and
using the chain rule, we get:
\[
mc\frac{\d T}{\d\tau}=\frac{\d E}{\d T}\frac{\d T}{\d\tau}=\frac{\d E}{\d\tau}=P=\varepsilon\sigma A\left(T^{4}-T_{\mathrm{ambient}}^{4}\right).
\]
Assuming that the temperature $T$ is much greater than the ambient
temperature, we may neglect $T_{\mathrm{ambient}}$ and rewrite this
relation as follows:
\[
\frac{1}{T^{4}}\d T=\frac{\varepsilon\sigma A}{mc}\d\tau.
\]
Integrating this equation from $T$ to the initial temperature\footnote{Note that $T<T_{0}$, so we must integrate from $T$ to $T_{0}$ in
order for the integral to be positive.} $T_{0}$, we get
\[
\frac{1}{3}\left(\frac{1}{T^{3}}-\frac{1}{T_{0}^{3}}\right)=\int_{T}^{T_{0}}\frac{1}{T^{4}}\d T=\frac{\varepsilon\sigma A}{mc}\int\d\tau=\frac{\varepsilon\sigma A}{mc}\tau.
\]
Isolating $T$, we find that the temperature of the object at time
$\tau$ is
\[
T\left(\tau\right)=\left(\frac{3\varepsilon\sigma A}{mc}\tau+\frac{1}{T_{0}^{3}}\right)^{-1/3}.
\]
Therefore, a massive object's temperature is a \textbf{monotonically
decreasing} function $T\left(\tau\right)$, where $\tau$ is the proper
time along a timelike geodesic.

It is interesting, although not crucial for our purposes, to allow
for massless ``objects'' as well. In this case, the geodesic is
lightlike, and $\tau$ is an affine parameter. We will assume that
the ``object'' is a photon gas, which satisfies the equation $N\sim VT^{3}$
where $N$ is the number of photons, $V$ is the volume, and $T$
is the temperature. The temperature is still monotonically decreasing,
as it is reduced by the gradual absorption of photons into atoms in
the air as the gas travels.

To prove that there is a paradox, we must first prove that the objects
always collide; if they don't, then there is no way for the object
to change its past trajectory, and thus there will be no paradoxes.
An object comes in radially from infinity towards decreasing values
of $r$, enters the time machine at time $t$, and then exits it at
$t-2$, now moving to infinity towards increasing values of $r$.
We thus have two half-infinite lines that are not parallel and therefore
\textbf{must cross}, so there will always be a collision.

Now, let's say that the object arrives in the vicinity of the time
machine at some initial temperature $T_{0}$. Since the temperature
is monotonically decreasing, after traversing a finite distance the
object enters the time machine at a strictly lower temperature $T_{1}<T_{0}$.
Since the throat length is zero, the object exits the time machine
in the past at the same temperature $T_{1}$. It moves a bit away
from the time machine until it reaches an even lower temperature $T_{2}<T_{1}$,
and then collides with its past self.

In this collision, the incoming (past) object will reverse its direction
of movement, so it is the time-traveling object that will go (again)
into the time machine. It will reach the time machine at some temperature
$T_{3}<T_{2}$, but since we already know that the object that left
the time machine had temperature $T_{1}$, for consistency we must
have $T_{3}=T_{1}$. This is impossible, since $T_{3}<T_{2}<T_{1}$and
all the inequalities are strict, so we have a consistency paradox.

Importantly, if the objects did not have temperatures, then we could
have applied Novikov's conjecture here, and argued that there is no
way to tell whether it is the future or past object that enters the
time machine, so there is no inconsistency.

The addition of temperature, much like the addition of color in the
old toy model, allows us to create a consistency paradox by forcing
an object to have two different temperatures at the same time. In
fact, the temperature could be replaced with any other measurable
property of the object that is monotonically decreasing or increasing,
and a similar inconsistency will be achieved.

\subsection{Bootstrap paradoxes}

Note that in addition to a consistency paradox, that is, a ``grandfather''-like
paradox, there is also a \textbf{bootstrap paradox }here. The object
that comes in from infinity never actually enters the time machine.
The other object comes out of the time machine and then ends up entering
the time machine again. Therefore, the object seems to be created
out of nothing, and only exists within the time loop.

One way to avoid a bootstrap paradox, while maintaining the consistency
paradox, is to create a situation where the two objects ``pass through''
each other instead of colliding. In the case of a massless photon
gas, this is already what happens -- therefore there is no bootstrap
paradox in that case.

To avoid a bootstrap paradox in the case of a massive object, we could,
in principle, consider a situation where the object is actually composed
of two disconnected pieces moving side by side in unison, so that
one pair of pieces can pass through the other pair without touching
it.

The temperature of the past pair will nevertheless decrease as a result
of ``passing through'' the future pair, as there will be some heat
exchange (however short) via radiation between the two pairs. Therefore,
there will still be a consistency paradox, since the object that will
enter the time machine will have a lower temperature than the object
that left the time machine.

To illustrate this, assume that a pair of pieces arrives in the vicinity
of the time machine at some temperature $T_{0}$. It moves towards
the entrance of the time machine, reaching some temperature $T_{1}<T_{0}$
somewhere along the path, and then enters the time machine at some
temperature $T_{2}<T_{1}$. It exits in the past at the same temperature
$T_{2}$. It moves a bit away from the time machine until it reaches
some temperature $T_{3}<T_{2}$, and then passes through the other
pair.

At the time of interaction, when the two pairs pass through each other,
the future (older) pair is at temperature $T_{3}<T_{2}$, and the
past (younger) pair is at temperature $T_{1}>T_{2}$. Therefore, the
past pair is the hotter one, and during the short interaction time,
it will transfer a small amount of heat to the colder future pair.
Thus, the past pair will now be at some temperature $T_{1}^{\prime}<T_{1}$,
and will then continue and enter the wormhole at some temperature
$T_{2}^{\prime}$.

We know that on the way from that particular point along the path
to the time machine, the pair originally decreased its temperature
from $T_{1}$ to $T_{2}$. Hence, regardless of the precise definition
of the monotonically decreasing function $T\left(\tau\right)$, since
the pair started at a temperature $T_{1}^{\prime}<T_{1}$ it must
enter the time machine at a temperature $T_{2}^{\prime}<T_{2}$.

But since we already know that the pair that left the time machine
had temperature $T_{2}$, we must have $T_{2}^{\prime}=T_{2}$, in
contradiction with the fact that $T_{2}^{\prime}<T_{2}$; even if
the temperature difference is extremely small, the temperatures are
still different. We thus still have a consistency paradox, but we
managed to avoid a bootstrap paradox.

This has been an interesting exercise, but unfortunately, it is a
very contrived way of resolving the bootstrap paradox, as it requires
delicate fine-tuning of the precise arrangement of the pieces, and
that the objects are composed of two pieces in the first place.

Since our stated goal in this paper is to avoid contrived situations
(such as the interaction vertices in our toy model), and instead get
comprehensive results from which we can learn about time travel paradoxes
in general, we do not consider the bootstrap paradox to be resolved,
except in this very special case -- at least not without introducing
multiple histories.

\subsection{Resolving the paradoxes using multiple histories}

We have proven that, with our wormhole time machine model, paradoxes
are inevitable, and \textbf{cannot} be resolved using Novikov's conjecture.
Consistency paradoxes are always created, and -- except in the special
contrived case where the objects can pass through each other -- bootstrap
paradoxes are always created as well. However, both types of paradoxes
\textbf{can} be resolved by introducing the notion of multiple histories
(or timelines), as we did with the toy model in \cite{ParadoxModel}.

We assume that the universe can have many histories, each distinguished
by a different value of a label $h$. Upon traveling back in time,
the universe ``branches'' into two separate histories. The original
history is left unchanged; the new history is the same up to the point
in time when the branching took place, but can be different from that
point on.

Any changes made to the new history will not influence the old history,
and thus no paradoxes can be created. We further impose that it is
impossible for the time traveler to go back from the new history to
the one they originally came from, with the possible exception of
going back to a point in time \textbf{after }they left, which does
not violate causality.

In this paper, we are not proposing a specific mechanism for creating
the new histories -- we are merely assuming that there is such a
mechanism already in place. The exact meaning of $h$, and the set
from which its values are taken (whether $\mathbb{N}$, $\mathbb{Z}$,
$\mathbb{R}$, or an even larger set) would depend on the particular
mechanism.

For example, if the multiple-history model is a consequence of the
Everett interpretation (see section \ref{subsec:The-nature-of}),
then different values of $h$ could represent different paths along
the branching tree of quantum possibilities.

The same -- as yet unknown -- mechanism that allows wormholes to
somehow connect two points in space and/or time, or even two different
universes, could presumably also be used to connect two different
histories.

Thus, in addition to the time shift $t\mapsto t-2$, it is a simple
matter to introduce a conjectured ``history shift'' $h\mapsto h+1$
(assuming, for simplicity, that $h$ is an integer). Note that this
does not\textbf{ }mean we are treating $h$ as a 5th spacetime dimension;
it is simply a label, and nothing more.

The reader is referred to \cite{ParadoxModel} for a discussion of
some interesting nuances of multiple-history models we did not cover
here, such as cyclic histories. Below, we will show that multiple
histories can resolve all of the paradoxes created by our model, both
consistency and bootstrap.

\section{\label{sec:Simulation-of-the}Simulation of the model using Mathematica}

The GitHub repository for this paper, at \href{https://github.com/bshoshany/WormholeParadoxSimulation}{https://github.com/bshoshany/WormholeParadoxSimulation},
contains the Mathematica notebook \texttt{WormholeParadoxSimulation.nb},
which simulates the wormhole time machine model presented in this
paper.

Note that purchasing Wolfram Mathematica is \textbf{not }required
to run the simulation; it can be viewed and interacted with using
Wolfram Player (version 12 and above), which can be freely downloaded
at \href{https://www.wolfram.com/player/}{https://www.wolfram.com/player/}.

Our time machine consists of the Morris-Thorne metric (\ref{eq:metric}),
with the wormhole at the spatial origin: $\left(t,l,\theta,\phi\right)=\left(t,0,\pi/2,0\right)$,
and with the two mouths separated in $t$ by 2 time units. Let the
object (whether a ball, a photon gas, or something else) begin at
$t=0$ in the initial position $l=l_{0}>0$. It then follows a radial
geodesic towards the wormhole, given by $l(t)=l_{0}-vt$ with a constant
velocity $v\in\left(0,1\right]$.

The object reaches $l=0$ at time $t=l_{0}/v$, at which point it
traverses the wormhole and continues to negative values of $l$. Equivalently,
since we are identifying the region of negative $l$ with the exact
same spacetime, except 2 units of time in the past, we can simply
interpret this as switching the sign of $v$. The object will exit
at the origin, with a time shift $t\mapsto t-2$, that is, at time
\[
t=\frac{l_{0}}{v}-2.
\]
The new worldline will thus be given by 
\[
l\left(t\right)=v\left(t-\left(\frac{l_{0}}{v}-2\right)\right)=v\left(t+2\right)-l_{0}.
\]
The old and the new worldlines will intersect when
\[
\underbrace{l_{0}-vt}_{\textrm{old}}=\underbrace{v\left(t+2\right)-l_{0}}_{\textrm{new}},
\]
that is, at time
\[
t=\frac{l_{0}}{v}-1.
\]
The collision will \textbf{always }happen; there is no way for the
new worldline to go ``around'' the old one, or to never reach it
in the first place, since they are both in the same plane and can
be extended to infinity.

\begin{figure}[h]
\begin{centering}
\includegraphics[width=1\textwidth]{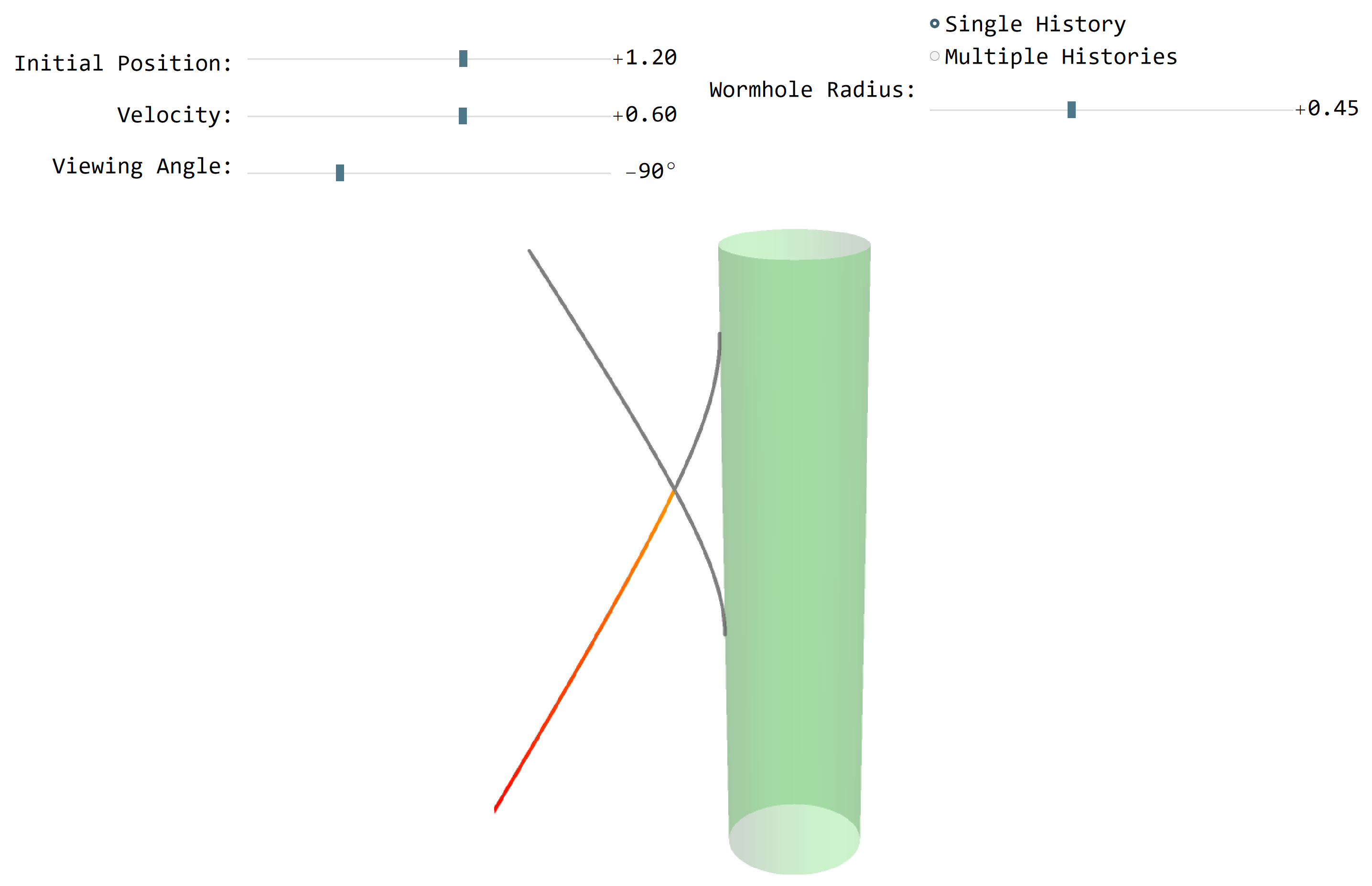}
\par\end{centering}
\caption{\label{fig:interface-persistent-single}The user interface of \texttt{WormholeParadoxSimulation.nb}
showing a wormhole with a single history. Note the gray lines, indicating
a paradox. The user may use the sliders to adjust the initial position
of the object, its velocity (in units of the speed of light), the
viewing angle of the plot, and the wormhole radius.}
\end{figure}

This scenario is simulated in our Mathematica notebook, as shown in
Figure \ref{fig:interface-persistent-single}. The wormhole is represented
in the figure by the green cylinder. Shown is a slice of the Morris-Thorne
spacetime with time $t$ as the vertical axis, the radial coordinate
$r\in\left[b_{0},\infty\right)$ orthogonal to the $t$ axis, and
the azimuthal angle $\phi$ going around the $t$ axis.

The polar angle $\theta$ is suppressed, such that\textbf{ each circle
in the cylinder represents a sphere} at a particular instance in time.
Note that in the Morris-Thorne spacetime, the region $r\in\left[0,b_{0}\right)$
-- the interior of the cylinder in the plot -- is inaccessible.

The incoming and outgoing paths of the object are also shown in the
figure, as calculated using the geodesic equations. Note that the
geodesics are shown in the $r$ coordinate, defined by $r^{2}\equiv b_{0}^{2}+l^{2}$,
instead of the $l$ coordinate -- which is why they are not straight
lines. The colors along the paths indicate temperature, with red being
the hottest.

The object starts at the red temperature and approaches the wormhole.
Upon traversal, the object exits the wormhole in the past, in the
opposite spatial direction, and then collides with its past self.
In the figure, we can see that some of the path is colored \textbf{gray},
which means there is \textbf{no consistent choice of temperature }along
the path, hence there is a paradox.

\begin{figure}[h]
\begin{centering}
\includegraphics[width=1\textwidth]{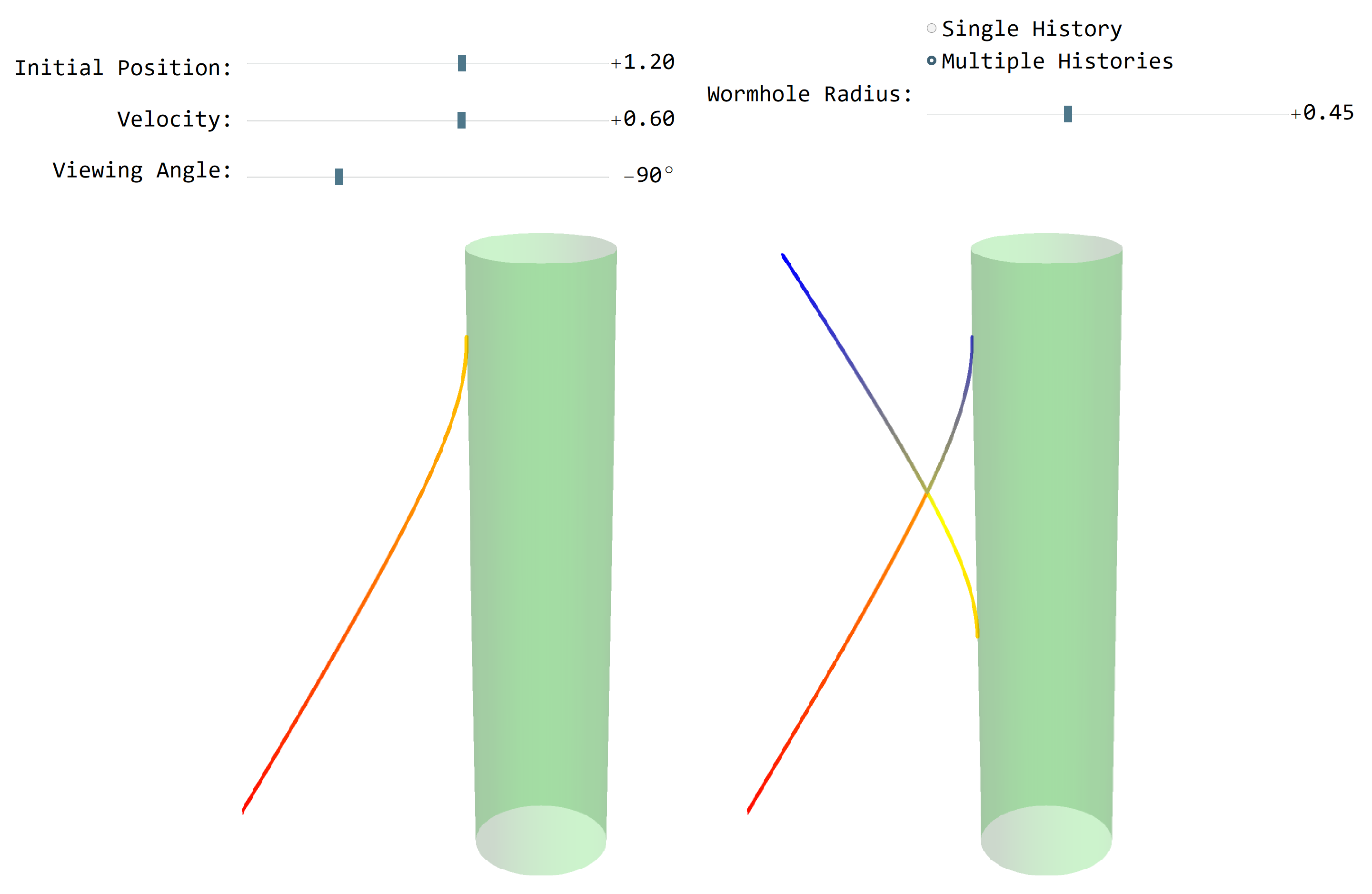}
\par\end{centering}
\caption{\label{fig:interface-persistent-multi}The user interface of \texttt{WormholeParadoxSimulation.nb}
showing a wormhole with multiple histories. Note that the paradox
is now resolved.}
\end{figure}

By switching from a single history to multiple histories, the paradox
can be resolved. Two histories will be shown on the screen, with the
first history on the left and the second (new) history on the right,
as in Figure \ref{fig:interface-persistent-multi}.

Notice that now there are no gray lines; there is a fully consistent
temperature evolution along the paths, and thus we have resolved the
consistency paradox. Furthermore, the object that emerges from the
time machine in the new history came from the old history, so it was
not ``created from nothing'', and thus the bootstrap paradox is
also resolved, without having to assume the objects pass through each
other.

\section{Additional arguments against Novikov's conjecture}

\subsection{\label{subsec:switch-paradox}Consistency paradoxes: the switch paradox}

In this paper we took great care to formulate a precise mathematical
model of a paradox and prove that Novikov's self-consistency conjecture
does not apply to it. For completion, we would now like to present
some other arguments which are more philosophical in nature, but nonetheless
worth mentioning.

Imagine that the wormhole time machine has an on/off switch. Upon
setting the switch to ``on'', matter is shifted in space such that
the spacetime geometry is that of a wormhole. Upon setting the switch
to ``off'', matter is shifted back to its original position, and
the wormhole disappears.

As above, the wormhole entrance is located 2 units of time in the
future of the wormhole exit. Alice intends to create a paradox using
the switch in the following way:
\begin{itemize}
\item At time $t=+1$, Alice goes into the wormhole entrance.
\item At time $t=-1$, Alice emerges from the wormhole exit and turns off
the switch, preventing herself from going into the wormhole at $t=+1$,
thus creating an inconsistency.
\end{itemize}
Assume that Alice carefully watches the switch from $t=-1$ until
$t=+1$. During that time, does she see her future self appear and
turn off the switch, or not?

If Alice never sees her future self, and then goes into the time machine
at $t=+1$, then we must be in a multiple-history scenario and Alice
must have been in the first history, ``before'' she traveled through
the time machine. There is no self-consistent chain of events where
Alice turns off the switch at $t=-1$ without her past self seeing
it happening, so Novikov self-consistency cannot apply.

If Alice does see her future self, but her future self does not turn
off the switch, then one must ask why she did not turn it off. Clearly,
not turning off the switch will avoid the paradox and create a fully
consistent chain of events. But we said above that Alice is determined
to turn off the switch in an attempt to create a paradox, no matter
what.

If, despite her determination, she did not turn off the switch, then
that seems to imply that Alice has no free will. While this is, by
itself, perfectly consistent with all known laws of physics, some
people may be concerned that accepting Novikov's conjecture means
we must reject the notion of free will.

If Alice does see her future self, and her future self does turn off
the switch, then the only self-consistent scenario is that something
mysteriously goes wrong that is beyond her control, such as the switch
suddenly malfunctioning\footnote{In fact, Alice changing her mind can also fall into this category.
While she is determined to turn off the switch, there is always a
small probability that she changes her mind after all. The issue isn't
that she changed her mind, but rather that she \textbf{had} to change
her mind regardless of how low the probability for that should have
been. Treated this way, we can avoid involving the controversial notion
of free will in the discussion.}.

Assuming that the switch was carefully constructed to be as reliable
as possible, the probability of it malfunctioning is presumably very
low, and yet to ensure self-consistency, an event with a seemingly
very high probability (the switch successfully turning off) must never
happen, and an event with a seemingly very low probability (the switch
malfunctioning) must happen with certainty.

This means that a selection process must be happening behind the scenes.
The local laws of physics, which were used to calculate the probabilities
of these events, must somehow be modified by an unseen hand to make
things consistent globally, selecting some events over others in a
way that would not make sense to a local observer unaware of the existence
of the time machine.

Indeed, one way to formulate Novikov's principle of self-consistency
is that ``the only solutions to the laws of physics that can occur
locally are those which are globally self-consistent'' \cite{PhysRevD.42.1915}.
This seems to contradict the local nature of the known laws of physics.

\subsection{\label{subsec:Bootstrap-paradoxes}Bootstrap paradoxes: the password
paradox}

As a final argument, we note that while Novikov self-consistency can
be used to resolve some consistency paradoxes, it cannot be used to
resolve bootstrap paradoxes. The reason is obvious from simply noting
the name \textbf{self-consistency}: this principle, by definition,
only guarantees consistency. But there are many perfectly consistent
scenarios that nonetheless suffer from bootstrap paradoxes -- where
something is created from nothing, or an event is its own cause.

Some may argue that bootstrap paradoxes are not really paradoxes,
and are instead ``pseudo-paradoxes'', as there is no reason for
the paradoxical chain of events to happen in the first place. However,
we will now argue that under certain conditions, Novikov's self-consistency
principle may guarantee that such a chain of events happens.

We turn the switch paradox into a bootstrap paradox as follows. The
switch starts at $t=-1$ in the on position, and at $t=0$, it turns
off automatically. In order to turn it back on, a secret password
must be entered. This password is \textbf{not} known to Alice. Furthermore,
for reasons that will become clear shortly, we assume that if the
switch is activated with the correct password, the person who activated
it is automatically (and inescapably) killed by the switch mechanism.

This seems to inevitably lead to a consistency paradox. Since Alice
doesn't know the password, she will not be able to re-open the wormhole.
But if the wormhole entrance is not open at $t=+1$, this is inconsistent
with the fact that the wormhole exit was open at $t=-1$.

This inconsistency can be solved using Novikov self-consistency as
follows. At $t=-1$, future Alice emerges from the wormhole exit.
At $t=0$, after the switch turns itself off, future Alice enters
the password ``42'', and activates the switch, which then kills
her instantly and opens the wormhole entrance.

With the wormhole entrance now open, past Alice enters it at $t=+1$,
and returns to $t=-1$. Knowing now that the password is ``42'',
she enters it and activates the switch. We have resolved the consistency
paradox, but in the process, we have created a bootstrap paradox --
Alice could not have known that the password is ``42'' had she not
seen herself type it in the first place.

What made this bootstrap paradox happen? One could, perhaps, reason
as follows. The local laws of physics allow, with extremely low probability,
for an object to spontaneously emerge out of the wormhole exit at
$t=-1$ due to random quantum fluctuations. There are a potentially
infinite number of possible objects that can be created in this way,
but only one of them is \textbf{globally self-consistent}: an Alice
with the knowledge that the password is ``42''.

Indeed, if anyone else comes out of the wormhole with the knowledge
that the password is ``42'', then this is inconsistent, because
once they enter the password, they die -- and only Alice remains
in the room, so only she can enter the wormhole at $t=+1$, and therefore
only she can exit it at $t=-1$.

In all other cases -- either Alice or someone else appears at the
wormhole exit without knowledge of the password, or some random object
appears, or even no object appears at all -- the wormhole entrance
will remain closed, and this is again inconsistent, since the exit
cannot exist without the entrance. 

While the event where Alice spontaneously appears at the wormhole
exit knowing that the password is ``42'' obviously has a very low
probability to happen, we have already established in the previous
section that Novikov self-consistency has the ability to make such
low-probability events happen with certainty in order to guarantee
consistency. Hence, we conclude that the self-consistency principle
will inevitably lead to a bootstrap paradox in the process of resolving
the password paradox.

\section{Conclusions and discussion}

\subsection{Our main argument}

Let us now summarize the main argument of our paper. We considered
a universe where time travel is possible, and in particular, it is
possible via Morris-Thorne wormholes. In this universe, we assumed
that the only two potential methods of resolving time travel paradoxes
are Novikov's self-consistency conjecture and multiple histories.

By proving that Novikov's self-consistency conjecture cannot be applied
to certain time travel paradoxes, we see that while it may well be
applicable to \textbf{some }time travel paradoxes, it cannot apply
to \textbf{all }paradoxes, and therefore, if time travel is indeed
possible, Novikov's conjecture alone is not sufficient to allow time
travel without any paradoxes.

Since we cannot simply sweep these Novikov-incompatible paradoxes
under the carpet, we conclude that multiple histories (or parallel
timelines) are a mandatory feature of any universe where time travel
is possible.

\subsection{Are our assumptions valid?}

It is worthwhile to make some comments regarding the validity of our
assumptions. Of course, the most radical assumption is that time travel
is possible. We do not yet have enough data to determine whether this
assumption is valid, and it may take many decades to get anywhere
near a concrete answer.

While some exotic metrics in general relativity seem to, at least
in principle, allow for the existence of closed causal curves, time
travel may still be forbidden, for example by some version of Hawking's
chronology protection conjecture \cite{Hawking92,visser1993wormhole,visser2003quantum},
as discussed in more detail in section \ref{subsec:Time-travel}.

However, this simply does not concern us here, as we are merely interested
in proving that multiple histories must exist \textbf{\uline{if}}\textbf{
}time travel is possible; we make no claims whatsoever regarding the
existence of time travel itself.

Even if time travel is indeed possible in our universe, the assumption
that it is possible via Morris-Thorne wormholes may still be incorrect.
First of all, we do not know if wormholes themselves can exist in
nature. It is often claimed that, since wormholes violate the energy
conditions \cite{Shoshany_FTL_TT,Curiel:2014zba}, they are unrealistic,
and cannot exist in our universe.

However, the energy conditions are seemingly arbitrary conditions
imposed by hand, and there are many known examples of both hypothetical
and real forms of matter which may violate them \cite{Lobo}. Still,
if some day it is proven -- from first principles -- that the violations
of the energy conditions required to create a wormhole are unrealistic,
then this would be enough to rule wormholes out. Furthermore, even
if wormholes are in fact realistic, at least given sufficiently advanced
technology, this does not by itself guarantee that they can be used
for time travel.

However, since in our proof we did not use any properties that are
unique to this particular spacetime, our argument should still stands
even if time travel is achieved by other means -- as long as the
objects that travel through time can have temperature and the initial
conditions can be set up such that collisions are inevitable.

Finally, there could be other potential methods of resolving time
travel paradoxes besides Novikov self-consistency and multiple histories\footnote{Interestingly, in \cite{ParadoxModel} we showed that in the case
of cyclic histories (there is a ``last'' history, and it connects
back to the first) one may combine Novikov's conjecture and multiple
histories into a ``hybrid'' method; but this is a special case that
will not be relevant to our discussion.}. However, to our knowledge, no other methods have been suggested
so far in the literature.

\subsection{\label{subsec:The-nature-of}Future plans}

If multiple histories must indeed exist, there must be some mechanism
for creating them and physical laws governing their existence. Our
goal in this paper was only to show that multiple histories are a
necessary consequence of time travel, but we make no claims regarding
the nature of these histories.

Although the possibility of resolving time travel paradoxes using
multiple histories has been occasionally mentioned before in the literature
\cite{Shoshany_FTL_TT,visser,Lobo}, an actual mechanism for creating
them has never, to our knowledge, been suggested. Developing such
a mechanism will not only provide concrete proof for the theoretical
possibility of paradox-free time travel; it will also allow us to
better understand the nature of time and causality in our universe.

Classically, non-Hausdorff manifolds \cite{visser,placek2014branching,pittphilsci16172,luc2019generalised}
or non-locally-Euclidean manifolds \cite{McCabe} were suggested as
a possible underlying mathematical model for branching spacetimes.
However, these models have many unresolved issues, mostly regarding
how to do differential geometry at points where the Hausdorff or locally-Euclidean
conditions are violated. Furthermore, no concrete model has so far
been suggested for relating such manifolds to time travel.

However, in quantum mechanics, the Everett (``many-worlds'') interpretation
seems to provide a natural and ``built-in'' way to account for multiple
histories. Deutsch \cite{Deutsch91} famously considered quantum mechanics
in the vicinity of closed causal curves and found that quantum mechanics
itself must be modified in order to resolve paradoxes. However, in
an upcoming paper \cite{EntangledTimelines}, we will provide a general
framework for resolving time travel paradoxes using unmodified quantum
mechanics.

Quantum mechanics alone may not be sufficient, though. To illustrate
this, consider again the switch paradox described in section \ref{subsec:switch-paradox}.
In this scenario, the spacetime geometry changes when the switch is
turned on or off. To resolve this paradox using multiple histories,
one would need to allow the second history to have a different spacetime
geometry than the first.

This means that the quantum state of spacetime itself must be considered,
and that cannot be done properly until a consistent and experimentally-verified
quantum theory of gravity becomes available. In a future publication,
we plan to discuss this problem in more detail.

In addition, it would be interesting to construct models for time
travel paradoxes which do not involve wormholes. Such models may be
based on other proposed forms of faster-than-light travel, such as
warp drives \cite{Alcubierre94} or hyperspace \cite{Hyperspace}.

This would allow us to study time travel paradoxes in scenarios very
different from both the TDP space model and the wormhole model, and
ensure that our conclusions regarding the necessity of multiple histories
for resolving time travel paradoxes are universally valid.

\section{Acknowledgments}

J. W. would like to thank Alicia Savelli for helpful discussions and
support, and Carlo Rovelli for his advice and insightful discussions.
B. S. would like to thank Thomas A. Roman for his invaluable input.
We also thank the two anonymous reviewers for identifying several
issues with the original manuscript. This research was supported by
funding from the Brock University Match of Minds grant.

\bibliographystyle{unsrturl}
\phantomsection\addcontentsline{toc}{section}{\refname}\bibliography{WormholeTimeMachine}

\begin{thebibliography}{10}

\bibitem{Shoshany_FTL_TT}
Barak Shoshany.
\newblock {Lectures on Faster-than-Light Travel and Time Travel}.
\newblock {\em SciPost Phys. Lect. Notes}, 10:1, 2019.
\newblock \href {https://arxiv.org/abs/1907.04178} {\path{arXiv:1907.04178}},
  \href {https://doi.org/10.21468/SciPostPhysLectNotes.10}
  {\path{doi:10.21468/SciPostPhysLectNotes.10}}.

\bibitem{Krasnikov}
Serguei Krasnikov.
\newblock {\em {Back-in-Time and Faster-than-Light Travel in General
  Relativity}}.
\newblock Springer International Publishing, 2018.
\newblock \href {https://doi.org/10.1007/978-3-319-72754-7}
  {\path{doi:10.1007/978-3-319-72754-7}}.

\bibitem{Lobo}
Francisco S.~N. Lobo, editor.
\newblock {\em {Wormholes, Warp Drives and Energy Conditions}}.
\newblock Springer International Publishing, 2017.
\newblock \href {https://doi.org/10.1007/978-3-319-55182-1}
  {\path{doi:10.1007/978-3-319-55182-1}}.

\bibitem{EverettRoman}
A.~Everett and T.~Roman.
\newblock {\em {Time Travel and Warp Drives: A Scientific Guide to Shortcuts
  Through Time and Space}}.
\newblock University of Chicago Press, 2012.
\newblock URL:
  \url{https://press.uchicago.edu/ucp/books/book/chicago/T/bo8447256.html}.

\bibitem{MorrisThorne88}
Michael~S. Morris and Kip~S. Thorne.
\newblock {Wormholes in spacetime and their use for interstellar travel: A tool
  for teaching general relativity}.
\newblock {\em American Journal of Physics}, 56(5):395--412, may 1988.
\newblock URL:
  \url{http://www.cmp.caltech.edu/refael/league/thorne-morris.pdf}, \href
  {https://doi.org/10.1119/1.15620} {\path{doi:10.1119/1.15620}}.

\bibitem{visser}
Matt Visser.
\newblock {Lorentzian Wormholes - From Einstein to Hawking}.
\newblock {\em Springer}, 1996.
\newblock URL: \url{https://www.springer.com/gp/book/9781563966538}.

\bibitem{PhysRevLett.61.1446}
Michael~S. Morris, Kip~S. Thorne, and Ulvi Yurtsever.
\newblock Wormholes, time machines, and the weak energy condition.
\newblock {\em Phys. Rev. Lett.}, 61:1446--1449, Sep 1988.
\newblock URL: \url{https://link.aps.org/doi/10.1103/PhysRevLett.61.1446},
  \href {https://doi.org/10.1103/PhysRevLett.61.1446}
  {\path{doi:10.1103/PhysRevLett.61.1446}}.

\bibitem{Krasnikov02}
S.~Krasnikov.
\newblock {The Time travel paradox}.
\newblock {\em Phys. Rev.}, D65:064013, 2002.
\newblock \href {https://arxiv.org/abs/gr-qc/0109029}
  {\path{arXiv:gr-qc/0109029}}, \href
  {https://doi.org/10.1103/PhysRevD.65.064013}
  {\path{doi:10.1103/PhysRevD.65.064013}}.

\bibitem{Krasnikov97}
SV~Krasnikov.
\newblock Causality violation and paradoxes.
\newblock {\em Physical Review D}, 55(6):3427, 1997.
\newblock \href {https://doi.org/10.1103/PhysRevD.55.3427}
  {\path{doi:10.1103/PhysRevD.55.3427}}.

\bibitem{wasserman2018paradoxes}
R.~Wasserman.
\newblock {\em Paradoxes of Time Travel}.
\newblock Oxford University Press, 2018.
\newblock URL:
  \url{https://global.oup.com/academic/product/paradoxes-of-time-travel-9780198793335}.

\bibitem{Hawking92}
S.~W. Hawking.
\newblock Chronology protection conjecture.
\newblock {\em Phys. Rev. D}, 46:603--611, Jul 1992.
\newblock \href {https://doi.org/10.1103/PhysRevD.46.603}
  {\path{doi:10.1103/PhysRevD.46.603}}.

\bibitem{visser1993wormhole}
Matt Visser.
\newblock {From wormhole to time machine: Comments on Hawking's chronology
  protection conjecture}.
\newblock {\em Phys. Rev.}, D47:554--565, 1993.
\newblock \href {https://arxiv.org/abs/hep-th/9202090}
  {\path{arXiv:hep-th/9202090}}, \href
  {https://doi.org/10.1103/PhysRevD.47.554}
  {\path{doi:10.1103/PhysRevD.47.554}}.

\bibitem{visser2003quantum}
Matt Visser.
\newblock {The Quantum physics of chronology protection}.
\newblock In {\em {The future of theoretical physics and cosmology: Celebrating
  Stephen Hawking's 60th birthday. Proceedings, Workshop and Symposium,
  Cambridge, UK, January 7-10, 2002}}, pages 161--176, 2002.
\newblock \href {https://arxiv.org/abs/gr-qc/0204022}
  {\path{arXiv:gr-qc/0204022}}.

\bibitem{earman2009laws}
John Earman, Christopher Smeenk, and Christian W{\"u}thrich.
\newblock Do the laws of physics forbid the operation of time machines?
\newblock {\em Synthese}, 169(1):91--124, 2009.
\newblock \href {https://doi.org/10.1007/s11229-008-9338-2}
  {\path{doi:10.1007/s11229-008-9338-2}}.

\bibitem{KimThorne}
Sung-Won Kim and Kip~S. Thorne.
\newblock Do vacuum fluctuations prevent the creation of closed timelike
  curves?
\newblock {\em Phys. Rev. D}, 43:3929--3947, Jun 1991.
\newblock URL: \url{http://resolver.caltech.edu/CaltechAUTHORS:KIMprd91}, \href
  {https://doi.org/10.1103/PhysRevD.43.3929}
  {\path{doi:10.1103/PhysRevD.43.3929}}.

\bibitem{Krasnikov1996}
S.~V. Krasnikov.
\newblock Quantum stability of the time machine.
\newblock {\em Phys. Rev. D}, 54:7322--7327, Dec 1996.
\newblock \href {https://arxiv.org/abs/gr-qc/9508038}
  {\path{arXiv:gr-qc/9508038}}, \href
  {https://doi.org/10.1103/PhysRevD.54.7322}
  {\path{doi:10.1103/PhysRevD.54.7322}}.

\bibitem{kay1997quantum}
Bernard~S. Kay, Marek~J. Radzikowski, and Robert~M. Wald.
\newblock {Quantum field theory on space-times with a compactly generated
  Cauchy horizon}.
\newblock {\em Commun. Math. Phys.}, 183:533--556, 1997.
\newblock \href {https://arxiv.org/abs/gr-qc/9603012}
  {\path{arXiv:gr-qc/9603012}}, \href {https://doi.org/10.1007/s002200050042}
  {\path{doi:10.1007/s002200050042}}.

\bibitem{krasnikov1998quantum}
S.~Krasnikov.
\newblock {Quantum field theory and time machines}.
\newblock {\em Phys. Rev.}, D59:024010, 1999.
\newblock \href {https://arxiv.org/abs/gr-qc/9802008}
  {\path{arXiv:gr-qc/9802008}}, \href
  {https://doi.org/10.1103/PhysRevD.59.024010}
  {\path{doi:10.1103/PhysRevD.59.024010}}.

\bibitem{PhysRevD.42.1915}
John Friedman, Michael~S. Morris, Igor~D. Novikov, Fernando Echeverria, Gunnar
  Klinkhammer, Kip~S. Thorne, and Ulvi Yurtsever.
\newblock Cauchy problem in spacetimes with closed timelike curves.
\newblock {\em Phys. Rev. D}, 42:1915--1930, Sep 1990.
\newblock URL: \url{https://link.aps.org/doi/10.1103/PhysRevD.42.1915}, \href
  {https://doi.org/10.1103/PhysRevD.42.1915}
  {\path{doi:10.1103/PhysRevD.42.1915}}.

\bibitem{Consortium91}
Fernando Echeverria, Gunnar Klinkhammer, and Kip~S. Thorne.
\newblock {Billiard balls in wormhole spacetimes with closed timelike curves:
  Classical theory}.
\newblock {\em Phys. Rev. D}, 44:1077--1099, Aug 1991.
\newblock URL: \url{https://authors.library.caltech.edu/6469/}, \href
  {https://doi.org/10.1103/PhysRevD.44.1077}
  {\path{doi:10.1103/PhysRevD.44.1077}}.

\bibitem{ParadoxModel}
Jacob Hauser and Barak Shoshany.
\newblock Time travel paradoxes and multiple histories.
\newblock {\em Phys. Rev. D}, 102, Sep 2020.
\newblock \href {https://arxiv.org/abs/1911.11590} {\path{arXiv:1911.11590}},
  \href {https://doi.org/10.1103/PhysRevD.102.064062}
  {\path{doi:10.1103/PhysRevD.102.064062}}.

\bibitem{Everett:2004ka}
Allen Everett.
\newblock {Time travel paradoxes, path integrals, and the many worlds
  interpretation of quantum mechanics}.
\newblock {\em Phys. Rev. D}, 69:124023, 2004.
\newblock \href {https://arxiv.org/abs/gr-qc/0410035}
  {\path{arXiv:gr-qc/0410035}}, \href
  {https://doi.org/10.1103/PhysRevD.69.124023}
  {\path{doi:10.1103/PhysRevD.69.124023}}.

\bibitem{Deutsch91}
David Deutsch.
\newblock Quantum mechanics near closed timelike lines.
\newblock {\em Phys. Rev. D}, 44:3197--3217, Nov 1991.
\newblock \href {https://doi.org/10.1103/PhysRevD.44.3197}
  {\path{doi:10.1103/PhysRevD.44.3197}}.

\bibitem{deutsch1994quantum}
David Deutsch and Michael Lockwood.
\newblock The quantum physics of time travel.
\newblock {\em Scientific American}, 270(3):68--74, 1994.

\bibitem{WHYTE20196}
Chelsea Whyte.
\newblock Paradox-free time travel.
\newblock {\em New Scientist}, 244(3261):6, 2019.
\newblock URL:
  \url{https://www.sciencedirect.com/science/article/pii/S0262407919323930},
  \href {https://doi.org/10.1016/S0262-4079(19)32393-0}
  {\path{doi:10.1016/S0262-4079(19)32393-0}}.

\bibitem{VisserBarcelo}
Matt Visser and Carlos Barcelo.
\newblock {Energy Conditions And Their Cosmological Implications}.
\newblock In {\em COSMO-99}, pages 98--112. World Scientific, 2000.
\newblock \href {https://arxiv.org/abs/gr-qc/0001099}
  {\path{arXiv:gr-qc/0001099}}, \href
  {https://doi.org/10.1142/9789812792129_0014}
  {\path{doi:10.1142/9789812792129_0014}}.

\bibitem{Curiel:2014zba}
Erik Curiel.
\newblock {A Primer on Energy Conditions}.
\newblock In Dennis Lehmkuhl, Gregor Schiemann, and Erhard Scholz, editors,
  {\em Towards a Theory of Spacetime Theories}, pages 43--104. Springer New
  York, New York, NY, 2017.
\newblock \href {https://arxiv.org/abs/1405.0403} {\path{arXiv:1405.0403}},
  \href {https://doi.org/10.1007/978-1-4939-3210-8_3}
  {\path{doi:10.1007/978-1-4939-3210-8_3}}.

\bibitem{Armendariz-Picon:2002gjc}
C.~Armendariz-Picon.
\newblock {On a class of stable, traversable Lorentzian wormholes in classical
  general relativity}.
\newblock {\em Phys. Rev. D}, 65:104010, 2002.
\newblock \href {https://arxiv.org/abs/gr-qc/0201027}
  {\path{arXiv:gr-qc/0201027}}, \href
  {https://doi.org/10.1103/PhysRevD.65.104010}
  {\path{doi:10.1103/PhysRevD.65.104010}}.

\bibitem{Kuhfittig:2020fue}
Peter K.~F. Kuhfittig.
\newblock {A note on the stability of Morris-Thorne wormholes}.
\newblock {\em Fund. J. Mod. Phys.}, 14:23--31, 2020.
\newblock \href {https://arxiv.org/abs/2009.11179} {\path{arXiv:2009.11179}}.

\bibitem{placek2014branching}
Tomasz Placek.
\newblock Branching for general relativists.
\newblock In {\em Nuel Belnap on indeterminism and free action}, pages
  191--221. Springer, Cham, 2014.
\newblock \href {https://doi.org/10.1007/978-3-319-01754-9_10}
  {\path{doi:10.1007/978-3-319-01754-9_10}}.

\bibitem{pittphilsci16172}
Joanna Luc and Tomasz Placek.
\newblock Interpreting non-hausdorff (generalized) manifolds in general
  relativity, June 2019.
\newblock URL: \url{http://philsci-archive.pitt.edu/16172/}.

\bibitem{luc2019generalised}
Joanna Luc.
\newblock Generalised manifolds as basic objects of general relativity.
\newblock {\em Foundations of Physics}, pages 1--23, 2019.
\newblock \href {https://doi.org/10.1007/s10701-019-00292-w}
  {\path{doi:10.1007/s10701-019-00292-w}}.

\bibitem{McCabe}
Gordon McCabe.
\newblock {The Topology of Branching Universes}.
\newblock {\em Foundations of Physics Letters}, 18(7):665--676, nov 2005.
\newblock \href {https://arxiv.org/abs/gr-qc/0505150}
  {\path{arXiv:gr-qc/0505150}}, \href
  {https://doi.org/10.1007/s10702-005-1319-9}
  {\path{doi:10.1007/s10702-005-1319-9}}.

\bibitem{EntangledTimelines}
Barak Shoshany and Zipora Stober.
\newblock Time travel paradoxes and entangled timelines.
\newblock 2023.
\newblock URL: \url{https://arxiv.org/abs/2303.07635}, \href
  {https://doi.org/10.48550/arXiv.2303.07635}
  {\path{doi:10.48550/arXiv.2303.07635}}.

\bibitem{Alcubierre94}
Miguel Alcubierre.
\newblock The warp drive: hyper-fast travel within general relativity.
\newblock {\em Classical and Quantum Gravity}, 11(5):L73--L77, may 1994.
\newblock \href {https://arxiv.org/abs/gr-qc/0009013}
  {\path{arXiv:gr-qc/0009013}}, \href
  {https://doi.org/10.1088/0264-9381/11/5/001}
  {\path{doi:10.1088/0264-9381/11/5/001}}.

\bibitem{Hyperspace}
Barak Shoshany.
\newblock Faster-than-light travel through hyperspace.
\newblock 2023.
\newblock In preparation.

\end{thebibliography}

\end{document}